\documentclass[11pt,a4paper]{article}
\usepackage{jheppub}
\usepackage{amsthm}
\usepackage{graphicx}

\title{Distinguishing among Technicolor/Warped Scenarios in Dileptons}

\author[a]{Piyali Banerjee}
\affiliation[a]{Group of Particle Physics, University of Montreal, Montreal QC, Canada} 
\author[b]{, Adam Martin} 
\affiliation[b]{Theoretical Physics Department, Fermi National Accelerator Laboratory, Batavia IL, USA}
 \author[c]{, and Veronica Sanz}
\affiliation[c]{Department of Physics and Astronomy, York University, Toronto ON, Canada}
 
\abstract{
Models of dynamical electroweak symmetry breaking usually include new spin-1 resonances, whose couplings and masses have to satisfy electroweak precision tests. We propose to use dilepton searches to probe the underlying structure responsible for satisfying these. Using the invariant mass spectrum and charge asymmetry, we can determine the number, parity, and isospin of these resonances. We pick three models of strong/warped symmetry breaking, and show that each model produces specific features that reflect this underlying structure of electroweak symmetry breaking and cancellations.}


\pdfoutput=1

\newcommand{\beq}{\begin{eqnarray}}
\newcommand{\eeq}{\end{eqnarray}}

\newcommand{\bmp}{\noindent\begin{minipage}{16cm}}
\newcommand{\emp}{\end{minipage}\vskip 7mm} 

\usepackage{dcolumn}
\usepackage{bm}
\usepackage{bbm}
\usepackage{subfigure}
\usepackage{pxfonts}
\usepackage{epstopdf}

\theoremstyle{definition}

\theoremstyle{plain}

\usepackage[ margin=5pt, font=normalsize,labelfont=bf,justification=raggedright]{caption}

\usepackage{slashed}
\def\lsim{\mathrel{\rlap{\lower4pt\hbox{\hskip1pt$\sim$}}
    \raise1pt\hbox{$<$}}}                
\def\gsim{\mathrel{\rlap{\lower4pt\hbox{\hskip1pt$\sim$}}
    \raise1pt\hbox{$>$}}}                

\def\beq{\begin{equation}}
\def\eeq{\end{equation}}
\def\bea{\begin{eqnarray}}
\def\eea{\end{eqnarray}}
\def\bq{\begin{quote}}
\def\eq{\end{quote}}
\def\to{\rightarrow}
\begin{document}
\maketitle 
\section {Introduction}

The idea that the electroweak phase transition is driven by new strong dynamics is not a new one~\cite{TCorig}. Dynamics are responsible for other phase transitions such as confinement in of color interactions, or superconductivity at low temperature. Unfortunately, with strong interactions, one is faced with intractable computations, and predictions thus entail large errors unless they rely on symmetries. The fact that that the idea of dynamical electroweak symmetry breaking is still under investigation after over 30 years attests to it attractiveness: it is a physical mechanism that already occurs in Nature and is devoid of unstable hierarchies. 

Ironically, large uncertainties do not save strong EWSB from facing very serious experimental constraints. One can estimate the effect of the new sector to the electroweak gauge boson parameters as measured at LEP by considering a reduced set of parameters, $S$, $T$ and $U$~\cite{stu}.
The inevitable conclusion is that strong EWSB would have shown up at LEP as a deviation from Standard Model predictions, unless symmetries or specific dynamics are in place. 

Model-builders are accustomed to implementing symmetries in order to suppress deviations, and they have achieved this for $T$ and $U$. The third parameter, $S$, has not been tamed by similar approaches -- small $S$ seems to require special dynamics, our understanding of which is more tenuous. Our arsenal is essentially limited to two main dynamical assumptions: {\it walking} and {\it warping}.

In the  picture of walking~\cite{walking}, a large anomalous dimension could lead to a parametric suppression of $S$. In the walking scenario a nearly marginal and slightly relevant operator runs
slowly and becomes strong and breaks electroweak symmetry. This picture often assumes the theory is near an interacting fixed point. 
It is unclear that a large anomalous dimension is possible in the context of electroweak symmetry breaking, but this idea is currently under study using lattice~\cite{latticeLSD} and analytical~\cite{Rattazzi} methods.

Warping relies on a holographic approach to strong dynamics~\cite{holography}. The holographic correspondence is set between the four-dimensional (4D) strong dynamics of interest, and a five-dimensional (5D) perturbative theory.    
 Holography provides insight into suppression mechanisms which are not
described in terms of symmetries. 5D suppressions are thus dynamical ones.
Indeed, in holography the localization in an extra dimension is interpreted
in 4D as effects from the non-perturbative dynamics, where the renormalization
group evolution is encoded into the wave-functions along the extra-dimension.

Interestingly, warping can be implemented in a way that would describe the effect of walking.
While warping does not address the origin of  the walking behaviour as a lattice study could, it can predict consequences on other observables that the lattice study cannot, such as production cross sections and lifetimes, and it is computationally less demanding than lattice. The two techniques are thus complementary.

The reason 5D models can be predictive is as follows. If it were possible to find a localization of fields in the extra dimension which suppresses some undesirable operators, such as the $S$-parameter, one could correlate the
localization assumption with some other effects, such as the spectrum and decay rates. In other words, while
this technique does not offer insight on the mechanism or dynamics underlying a suppressed
operator, it does allow us to predict  observable consequences. The literature describes many implementations of this idea: approaches to describe solutions to the gauge~\cite{gauge} and mass hierarchies~\cite{hierarchies}
and flavor problems~\cite{flavor}, have all been addressed in the holographic picture as a consequence
of localization inside the bulk, and not as a consequence of symmetries.

Warped or walking, strong dynamics leads to scenarios where new resonances show up as composite objects of the strong dynamics. The common prediction to all those models is that the resonances would couple strongly to $W,Z$ bosons and help in the unitarization of $WW$ scattering. Unfortunately, experimental access to this prediction is very limited~\cite{VBF}. In this paper, we propose a different approach:  scenarios of strong dynamics may be distinguished using simple channels such as dileptons.  In fact, even if the resonance couplings to light fermions is suppressed, the s-channel production may turn out to be the discovery channel: this channel is very clean and provides charge correlations. 

Indeed, many scenarios predict sizable s-channel production. In this paper, we focus on three distinct scenarios based on warped extra-dimensions and technicolor scenarios, and use dileptons as the discovery and also characterization channel. The models considered here are Cured Higgsless (CHL), Holographic Technicolor (HTC) and Low-Scale Technicolor (LSTC), and we outlined their main characteristics in the text. The main point to take away is that each of these models addresses electroweak precision tests in a specific way, and that this information is encoded in the spectrum of resonances, and in their parity.

The paper is organized as follows: In section~\ref{sec:dileptons}, we relate the warping and walking scenarios to the mass reconstruction in the dilepton final state. In section~\ref{sec:dilepbound} we recast the current LHC bounds on dilepton resonances in terms of lepton-resonance couplings for each of the three models. Next, in section~\ref{sec:massreco} we perform a simulation of the di-electron mass reconstruction for each model. As an accurate characterization of the lepton resolution is crucial to determining how well nearby resonances can be distinguished, we model the detector response using the fast simulator ATLFAST++~\cite{atlfast-software}. Once the masses of new resonances have been determined, their coupling structure is the next question to answer. We discuss a simple, low-statistics method to address the couplings structure in section~\ref{sec:angdist}, then conclude with a discussion.

\section{Technicolor/Warped in Dileptons}
\label{sec:dileptons}

 \begin{figure}[t]
\centering
\includegraphics[scale=0.25]{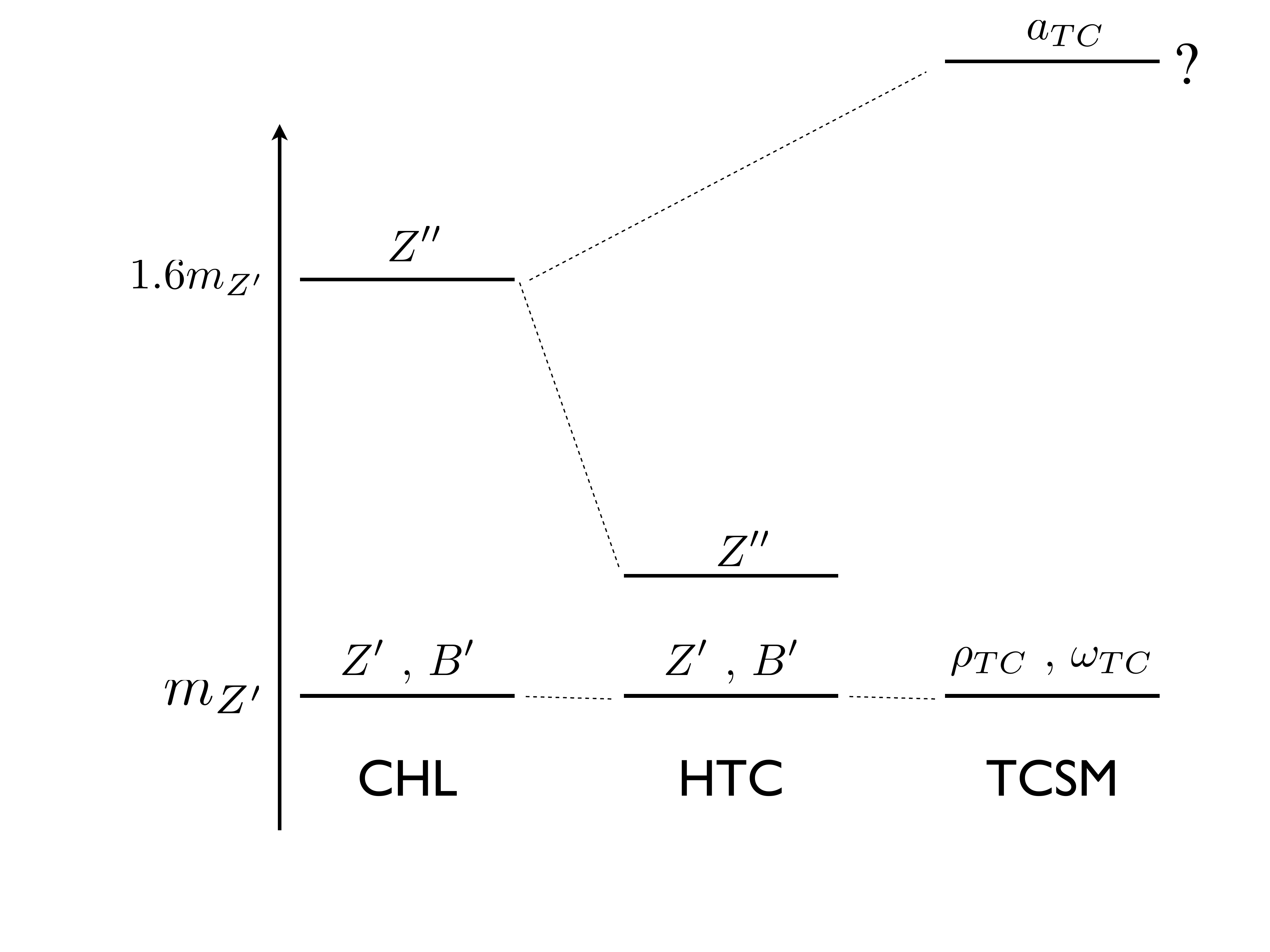}
\caption{The spectrum of dilepton resonances in the Cured Higgless (CHL), Holographic Technicolor (HTC) and Low Scale Technicolor (TCSM) cases. In TCSM, the splitting between the first and second tier of resonances is not specified.}
\label{cartoon}
\end{figure}  

In this paper we focus on three different models of dynamical electroweak symmetry breaking; two are five-dimensional and therefore are 'warped', while the third model is a 'walking', purely four-dimensional scenario.

 The holographic approach can be used to achieve $S\cong 0$, and there the suppression has
a definite bearing on the spectrum of a theory. Indeed, possible ways of obtaining a small S in
5D models involve either a direct modification of the spectrum of spin-1 fields (Holographic
Technicolor~\cite{HTC}, or HTC) or a balance of spin-1 versus spin-1/2 particles (Cured Higgsless~\cite{matching,higgsless}, or CHL).  In this paper, we are going to focus on characterizing strong EWSB using the dilepton final state. In HTC, the cancellation of spin-one resonances in the $S$ parameter requires two close-by resonances. Those two resonances would show up in dileptons, whereas in CHL there is only one low-lying resonance, and another resonance waits at a larger mass, about 1.6 times the mass of the first resonance. CHL is a model based on an Anti-de Sitter (AdS) geometry in 5D\cite{higgsless, deconstructed}, and the ratio 1.6 is just a ratio between the zeroes of Bessel functions~\footnote{Let us mention that in all these models one would expect excitations of the $B$ and $W^3$ gauge bosons. Yet, the splitting between those two states is very small, typically smaller than the experimental resolution, and has no consequences on the discussion in this paper.}. HTC is a model based on 5D warped space-time, but the geometry of HTC is no longer pure AdS, but AdS with large deviations from conformality\cite{AdSQCD}. Those deviations in the geometry are mapped to the presence of condensates breaking the conformal symmetry of  a strongly coupled sector. For example, in the language of QCD, those deviations from conformality correspond to quark and gluon condensates\cite{AdSQCD}. 

Our third scenario, Low-scale technicolor (LSTC) (also called the technicolor straw man, TCSM)~\cite{TCSM} takes a different approach. It gives up calculability, assumes walking, and is more phenomenologically driven, see Ref.~\cite{MWT} for a different popular four-dimensional scenario. Rather  than model the strong dynamics, LSTC introduces a small parameter in a sector of the theory. Some quantities can then be modeled and related to each other as an expansion in this small parameter. Given these model dependencies, one still expects a resonance to lie at low energies playing the role of a techni-rho, $\rho_{TC}$. The scale of the next resonance, the techni-axial ($a_{TC}$) is not fixed in this model, although one could take QCD as a guidance, where the ratio between the $\rho$ and $a_1$ is about 1.7, very close to the warped scenario. The spectrum of LSTC also differs from the holographic models in that it contains technipions $\pi_T$ -- uneaten pseudo-Goldstone bosons that are typically the lightest composite states in the spectrum (other than the $W/Z$) and which couple strongly to the spin-1 resonances. Technipions couple to SM quarks and leptons proportionally to the fermion's mass, thus they rarely decay to leptons. The only impact the technipions have on our study is indirect; if allowed, the $\rho_T, a_T$ prefer to decay to technipions, thereby changing the branching fraction of the $\rho_T, a_T$ to leptons.

In Fig.~\ref{cartoon} , we show the vector spectrum for CHL, HTC and TCSM. In CHL, the splitting between the first and second tier of resonances is set by the AdS geometry. In HTC, the splitting is set by the requirement that the first and second tier of resonances conspire with each other to cancel contributions in the $S$ parameter. As a consequence, in HTC the two tiers are close to each other, although the degeneracy can be resolved experimentally (see Sec.~\ref{sec:massreco}). Finally, TCSM makes no assumptions about the spectrum besides the presence of a vector resonance at low energies.

Notice that the above differences between models are not casual, but rather, reflect the deeper structure of each model, and the way electroweak cancellations are built into it. 
 
\section{Current bounds on dilepton resonances}
\label{sec:dilepbound}

Dilepton resonances are a clean search channel, and there is an ongoing improvement of these searches as colliders analyze more data. Obviously, the bounds coming from these searches depend on the resonance mass and its coupling to light quarks and leptons.
In this paper we are going to consider resonances with masses around the TeV, and the best limits for this mass range come from LHC.

The ATLAS and CMS collaborations are looking for new resonances into dileptons, and results with an integrated luminosity of about 1 fb$^{-1}$ are available~\cite{ATLASdilep, CMSdilep}. Both collaborations obtain similar results, but we are going to focus on the ATLAS limits because their results show a comparison with different models, including a sequential $Z'$ (a heavy spin-one resonance with couplings equal to the SM $Z$ boson). ATLAS quotes a bound on the cross section times branching ratio of 
\bea 
\sigma \times B \lesssim 10^{-2} \textrm{ pb}
\label{limits}
\eea
for a resonance mass of about 700 GeV.  Assuming that the acceptance of our $Z'$ to the cuts used for these search are the same as quoted for a sequential $Z'$ of the same mass, one can recast the limit in Eq.~\ref{limits} in a limit of coupling and branching ratio to leptons,
\bea
\left( \frac{g_{Z' f \bar{f}}}{ g_{SM}} \right)^2 \,  \frac{B_{mod}}{B_{SM}} < 0.014 \ ,
\label{limits2}
\eea
where $B_{mod}$ ($B_{SM}$) is the branching ratio of a new resonance (the $Z$) boson to dileptons (lepton=$e^{\pm}$, $\mu^{\pm}$). For example, if $B_{mod}=B_{SM}$, the bound on the coupling of the $Z'$ to light fermions can be rewritten as $g_{Z' f \bar{f}} < 0.12 \,   g_{SM}$. 

Among the three scenarios considered in this paper, the only model predicting a specific range of couplings to light fermions is CHL. The light fermion-resonance couplings in CHL are tied to the cancelation required in the $S$ parameter. Even within the context of the CHL model, the allowed range of couplings is fairly large for a resonance in the TeV range, $g_{Z' f \bar{f}} \lesssim  0.15$, see Tables 3 and 4 in Ref.~\cite{matching}. In the mass point we use in the paper, the branching fraction of the first resonance is $B_{CHL}\simeq 4 \times 10^{-3}$, which is about a factor 8 below the SM branching ratio to electrons.
 
In our second holographic model, HTC, electroweak precision measurements do not set strong constraints on the coupling of resonances to light fermions since the mechanism of canceling the contributions to the $S$ parameter involves uniquely the vector sector. Couplings in HTC are therefore a free parameter and can accommodate the LHC bounds. 

Our walking model does not address the problem of large contributions to the electroweak precision tests, and therefore there is no relation between the couplings of SM fermions to the new resonances coming from indirect measurements. Nevertheless, the tecnirho ($\rho_T$) branching ratio to electrons depends on the assumptions about the technipion ($\pi_T$) mass. Assuming $\rho_T \to \pi_T \pi_T$ is not allowed (as is usually done), the branching fraction $\rho_T \to  e^+e^-$ varies from 0.002 ($m_{\pi_T} \sim m_{\rho_T}$) to 0.009 ($m_{\pi_T} + m_{W/Z} > m_{\rho_T}$)~\cite{TCSM}.

In summary, all models are constrained by LHC searches in dileptons, but it does not yet impinge on the parameter space we are interested in this paper.

 \section{Mass reconstruction}
 \label{sec:massreco}
 
For all the three models CHL, HTC and LSTC we reconstruct the mass of the 
resonances in the setting of the ATLAS detector of the Large Hardon Collider
running at 7 TeV COM energy. For detector effects and 
reconstruction we use Atlas Fast Simulation Program ATLFAST++~\cite{atlfast-software}, 
which is a ROOT-based standalone C++ program. 

\begin{figure}[t] 	
\begin{center}
\includegraphics[width=5cm]{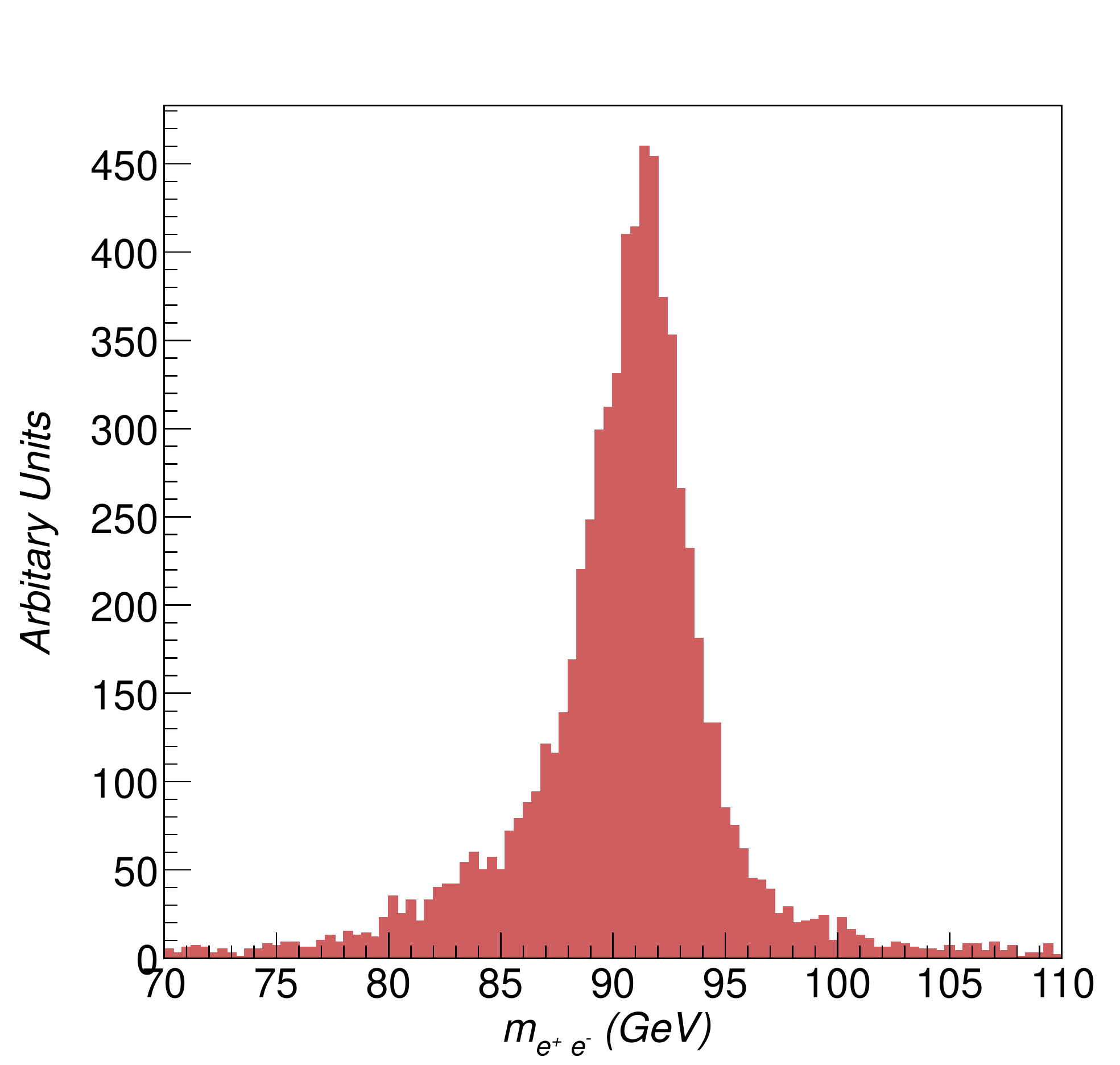} 
\includegraphics[width=5cm]{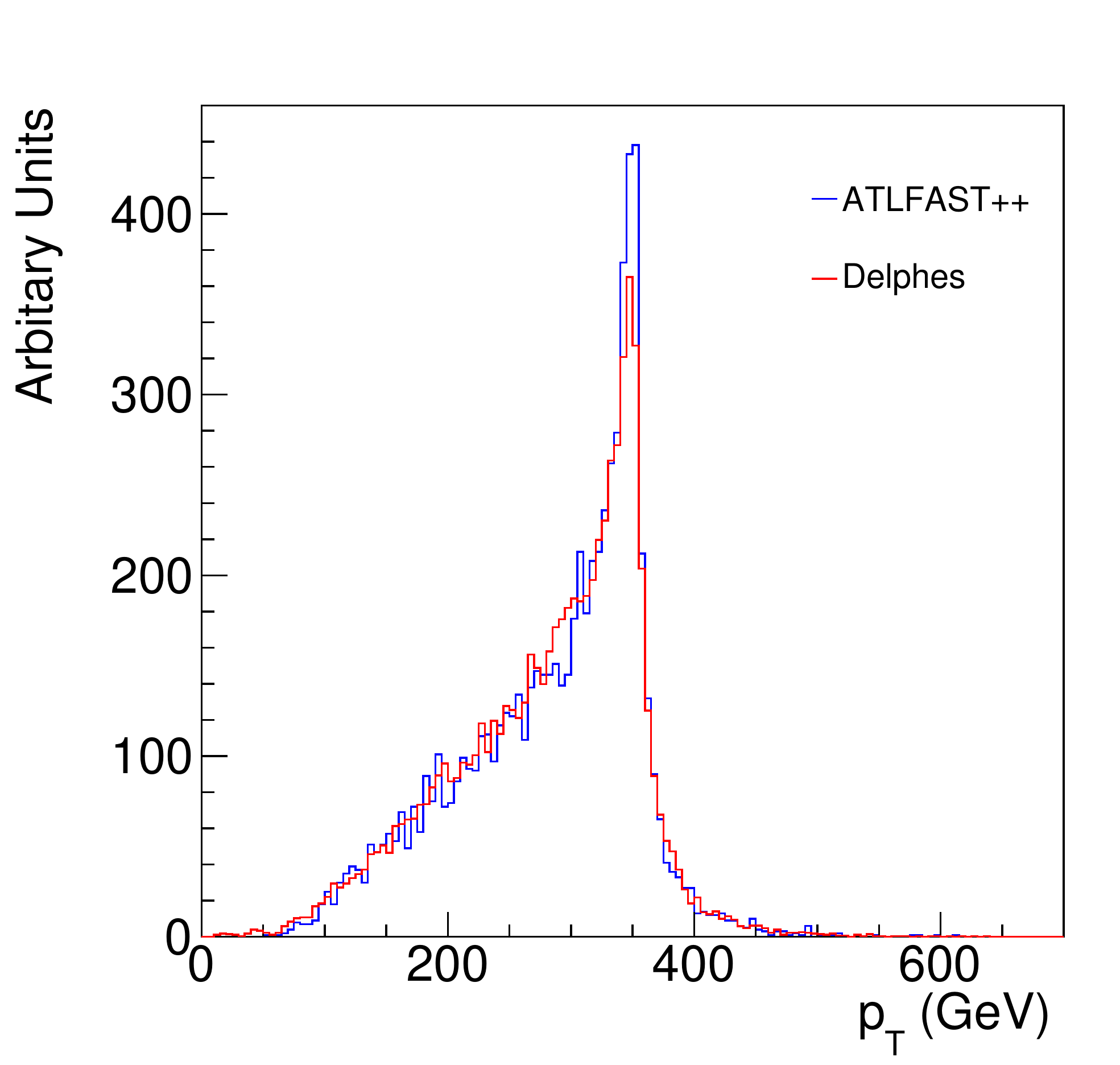} 
\includegraphics[width=5cm]{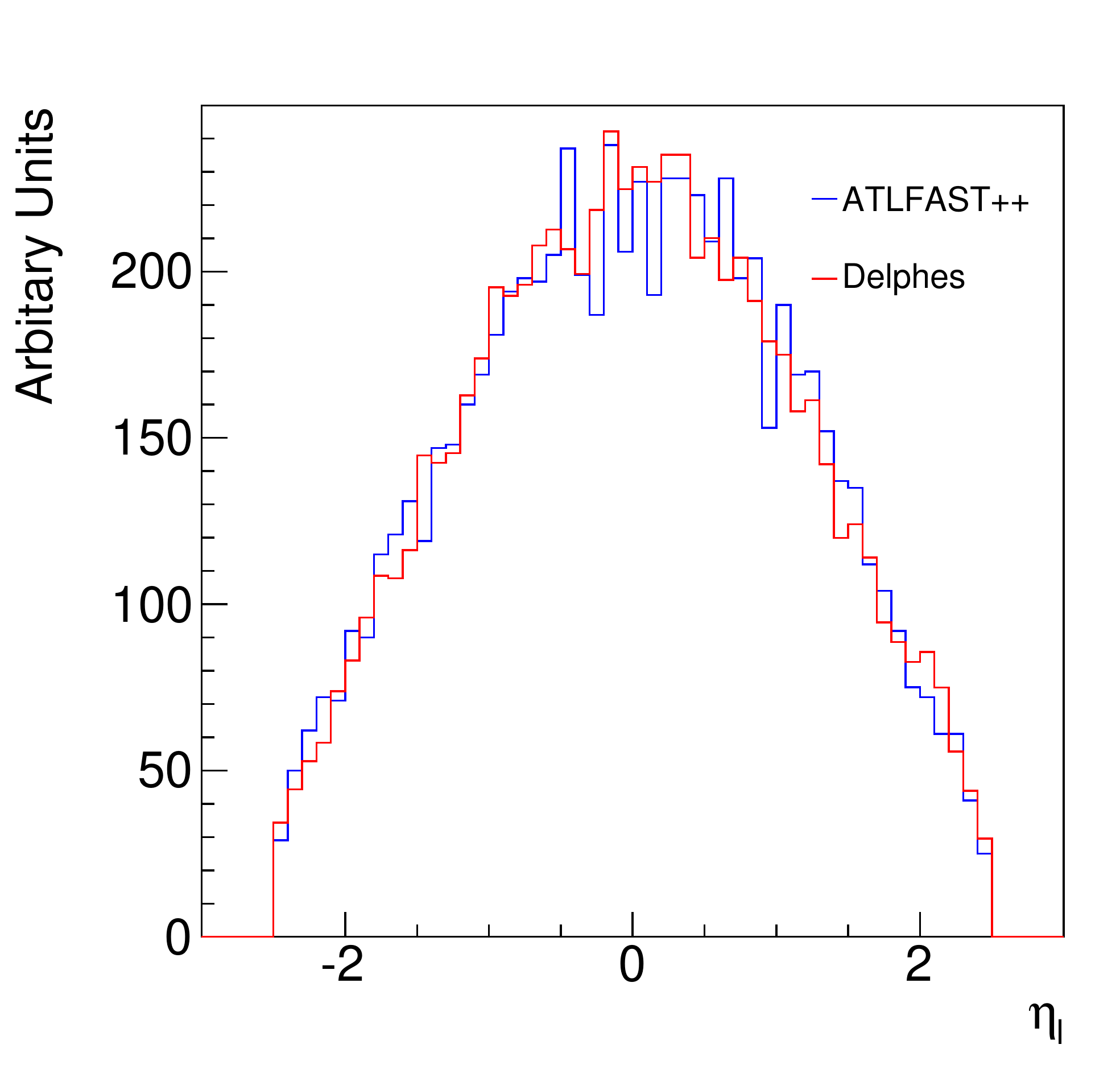} 
\end{center}
\caption{
Left: ATLAFAST++ dielectron mass reconstruction of a $Z$ boson. Right: Resolutions arising from ATLFAST++ and DELPHES. Electron $p_T$ and $\eta$ distributions. }
\label{fig:atlfastreso}
\end{figure}

Before showing our results for the three models, in  Fig.~\ref{fig:atlfastreso} we compare the lepton resolution in ATLFAST++ with full simulation in ATLAS~\cite{atlasreso} and another simplified detector simulator,  DELPHES~\cite{Ovyn:2009tx}.  
In the left panel, we plot the dielectron mass reconstruction of a $Z$ boson in ATLFAST++. The resolution in this channel agrees with the results from a full-simulation analysis reported by ATLAS~\cite{atlasreso}, and with the mass reconstruction using 2010 data, See Fig. 3 in~\cite{ATLASdilepnew}. 
 
While the agreement with~\cite{atlasreso} is encouraging,  our signal is not the SM $Z$ boson, but a heavy resonance which decays into high-$p_T$ leptons. Therefore, as a sanity check we use another standalone detector simulator, DELPHES, to compare with  ATLFAST++ results. On the right of Fig.~\ref{fig:atlfastreso} we show the $p_T$ and $\eta$ distributions of the leptons. The two simulations match, showing that the signal is characterized by central and high-$p_T$ leptons.  

Thorough studies have been performed on the resolution of electrons and muons coming from dilepton resonances. As seen inTable II of ATLAS note Ref.~\cite{ATLASdilep}, the electron channel usually leads to the best resolution. Therefore in this paper we will concentrate on results from the dielectron channel alone. We expect similar, though slightly weaker conclusions using the dimuon channel. 

The signal is generated using the Madgraph event generator~\cite{Alwall:2011uj} for each of 
the models with the lowest mass resonance set to 700 GeV. These generator events are 
then passed through Pythia~\cite{Sjostrand:2006za} for hadronization and parton showering. The events from Pythia are then passed through 
ATLFAST++ to simulate the detector effects.  We then sort 
the electrons in ascending order in $p_T$ and select the two highest $p_T$
electrons for our invariant mass reconstruction.
We apply a $p_T$ cut of 25 GeV or higher to the 
two electrons to reduce the background from QCD fakes.  
 
 \begin{figure}[t]
\begin{center}
\includegraphics[width=7.5cm]{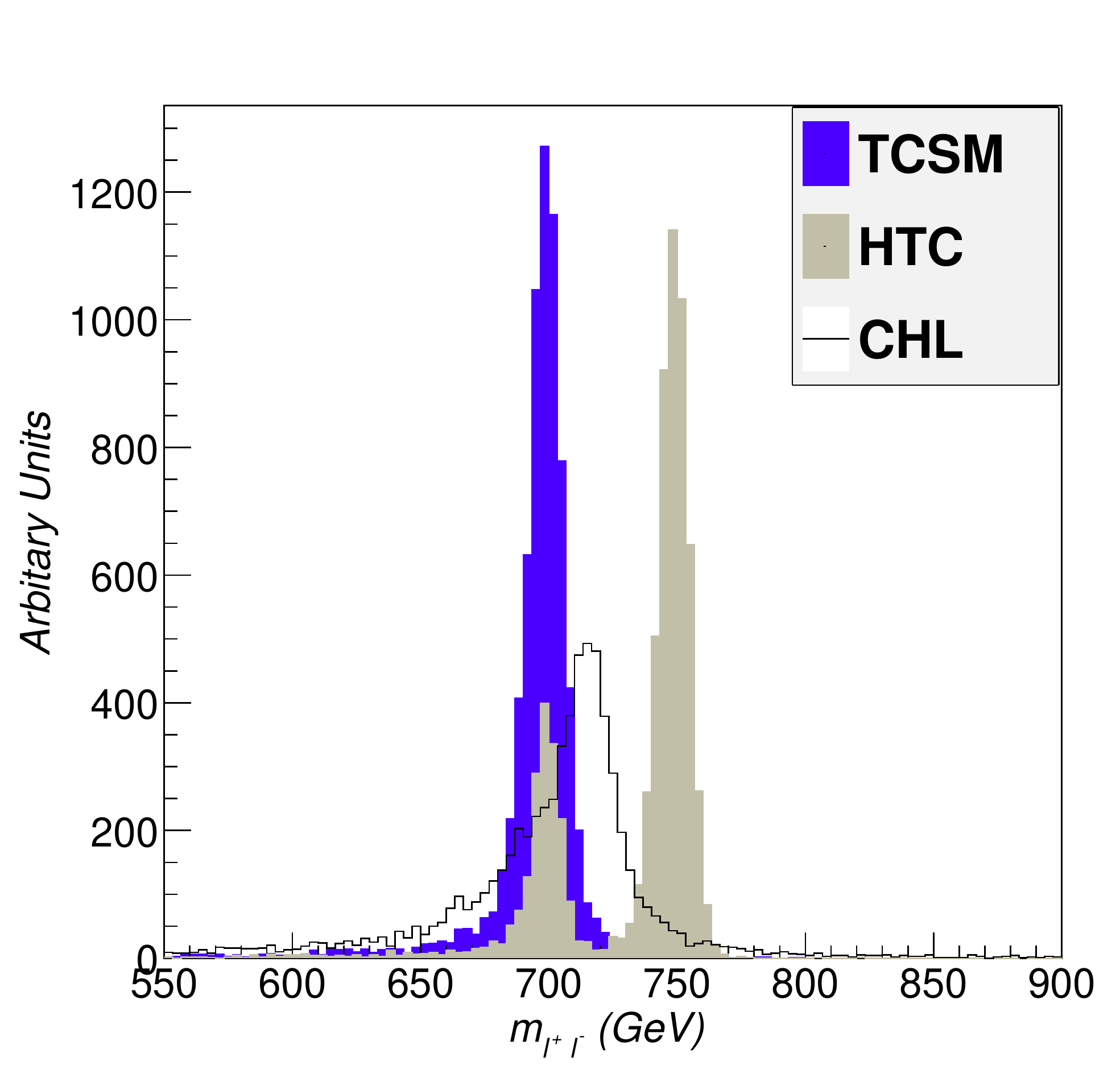} 
\includegraphics[width=7.5cm]{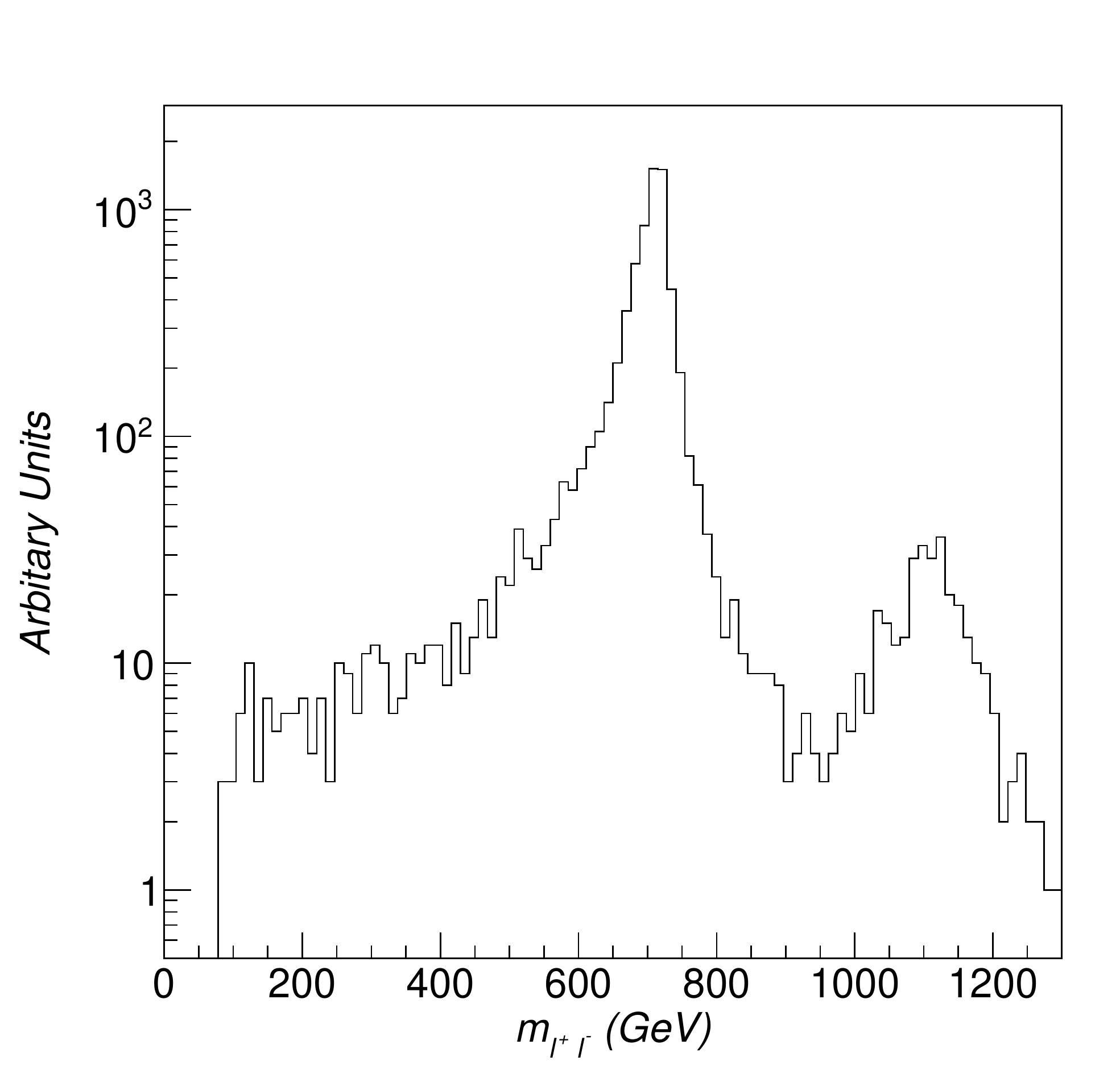} 
\end{center}
\caption{Left: Dilepton invariant mass for the three models. The dark blue histogram corresponds to TCSM, the nearby resonances in light grey are the HTC spectrum, and the black line is the first tier of resonances in CHL. Right: Dilepton invariant mass in CHL. Note the logarithmic scale.}
\label{fig:mass}
\end{figure}

The left side of figure~\ref{fig:mass} shows the mass of the lightest resonance for all the three models in the dielectron channel. 
The two nearby mass peaks for HTC are well separated within experimental resolution.
Note that the mass resolution for TCSM and HTC is dominated by experimental
effects while for CHL it is dominated by the theoretical prediction. In HTC, the resonances are comparatively broad due to enhanced decays to tops-- the coupling of the resonances to top quarks is large due to partial compositeness of the top. 

The nearby presence of two narrow resonances in HTC is a prediction tied to electroweak precision measurements~\cite{HTC}. This prediction is  easily tested in a clean dilepton channel. In CHL, a second tier of resonances should show up at $m_{2nd} \sim 1.6 m_{1st}$. In the right part of Fig.~\ref{fig:mass} we show a larger mass range in the dileptons, where the second tier is visible. Obviously, the discovery of this second resonance would require a larger luminosity.

Regarding backgrounds, the main contribution comes from SM Drell-Yan processes with an intermediate off-shell photon or $Z$ boson. Smaller contributions come from $t\bar{t}$, dibosons and cosmic rays. Those have been studied in  ~\cite{ATLASdilep, CMSdilep}, and taken into account when drawing a limit on the total cross section in Eq.~\ref{limits}. To reduce the backgrounds from DY processes, one would apply an invariant mass and lepton $p_T$. In this paper we do not simulate the DY backgrounds in the signal region ($m_{e^+e^-}\simeq 700$ GeV) because we are not setting the overall normalization of the signal, but rather assume the signal total production would be below the current limit.
  
 \section{Using the charge asymmetry to probe vector meson dominance}
 \label{sec:angdist}
 
Dilepton final states provide an excellent energy-momentum resolution, but also a precise charge identification. Therefore, besides the mass spectrum, one can obtain a rather accurate measurement of the lepton charge asymmetry in the events. As we will detail below, this asymmetry provides a measurement of the chirality of the resonance coupling to light fermions. 

The measurement of the chirality of couplings to fermions is a test of the assumption of {\it  vector meson dominance} (VMD)~\cite{VMD}, often used in models of strong electroweak symmetry breaking. In a vector meson dominance scenario the first resonance: {\it 1.)} has vector couplings to fermions (before electroweak symmetry breaking), and {\it 2.)} is well separated from the next tiers. CHL and TCSM are models of VMD, whereas HTC addresses the electroweak precision tests problem by largely differing from VMD.  Also note that after electroweak symmetry breaking, the resonances mix with the electroweak gauge bosons, which further modifies the chirality of the heavy mass eigenstate. Therefore, for all the models here, measuring the chirality of the coupling is a combined measurement of the VMD assumption and the mixing with the SM electroweak gauge bosons.

To obtain an expression for the asymmetry, let us write the couplings of the resonance to fermions as
 \bea
 Z'_{\mu} \bar{f} \gamma^{\mu} (V+\gamma_5 A) f
 \eea
 where $Z'$ is a new resonance, $V=\frac{L+R}{\sqrt{2}}$ and $A=\frac{L-R}{\sqrt{2}}$ are the axial and vector couplings, and $L$, $R$ are the couplings to left-handed (LH) and right-handed (RH) fermions.

As we mentioned before, the chiral structure  of this coupling is especially important in HTC, where one expects the two nearby resonances to be an admixture of vector an axial interaction states. In CHL and TCSM, one also expects an admixture of $V,A$ couplings of $Z'$ to fermions, but the admixture is purely due to mixing with the SM $Z$ boson, and is therefore suppressed as $(m_{Z}/m_{Z'})^2$~\cite{mixingSM}. 

To gain information on the chirality of the couplings, we first focus on the parton level process 
 \bea
 q(p) \bar{q} (p') \to Z' \to \ell^- (k) \ell^+ (k') \ .
 \eea

At the $Z'$ pole, one would write
 \bea
 A_{FB}=\frac{F-B}{F+B} =\frac{3}{4} \, A_{\ell} A_{q} 
 \label{partonAFB}
 \eea
 where 
 \bea
 F & = & \int_0^1 d \, cos \theta \frac{d \sigma}{d \theta} \\
 B & = & \int_{-1}^0 d \, cos \theta \frac{d \sigma}{d \theta}
 \eea
and
 \bea
 A_{f} \propto \frac{A.V}{A^2+V^2} \ ,
 \label{coeff}
 \eea
 where $f=q,\ell$.
Obviously, this asymmetry vanishes in the pure vector ($A=0$) and pure axial ($V=0$) cases. This conclusion holds beyond the parton level as it only depends on the coupling.

\begin{figure}[t]
\centering
\includegraphics[scale=0.2]{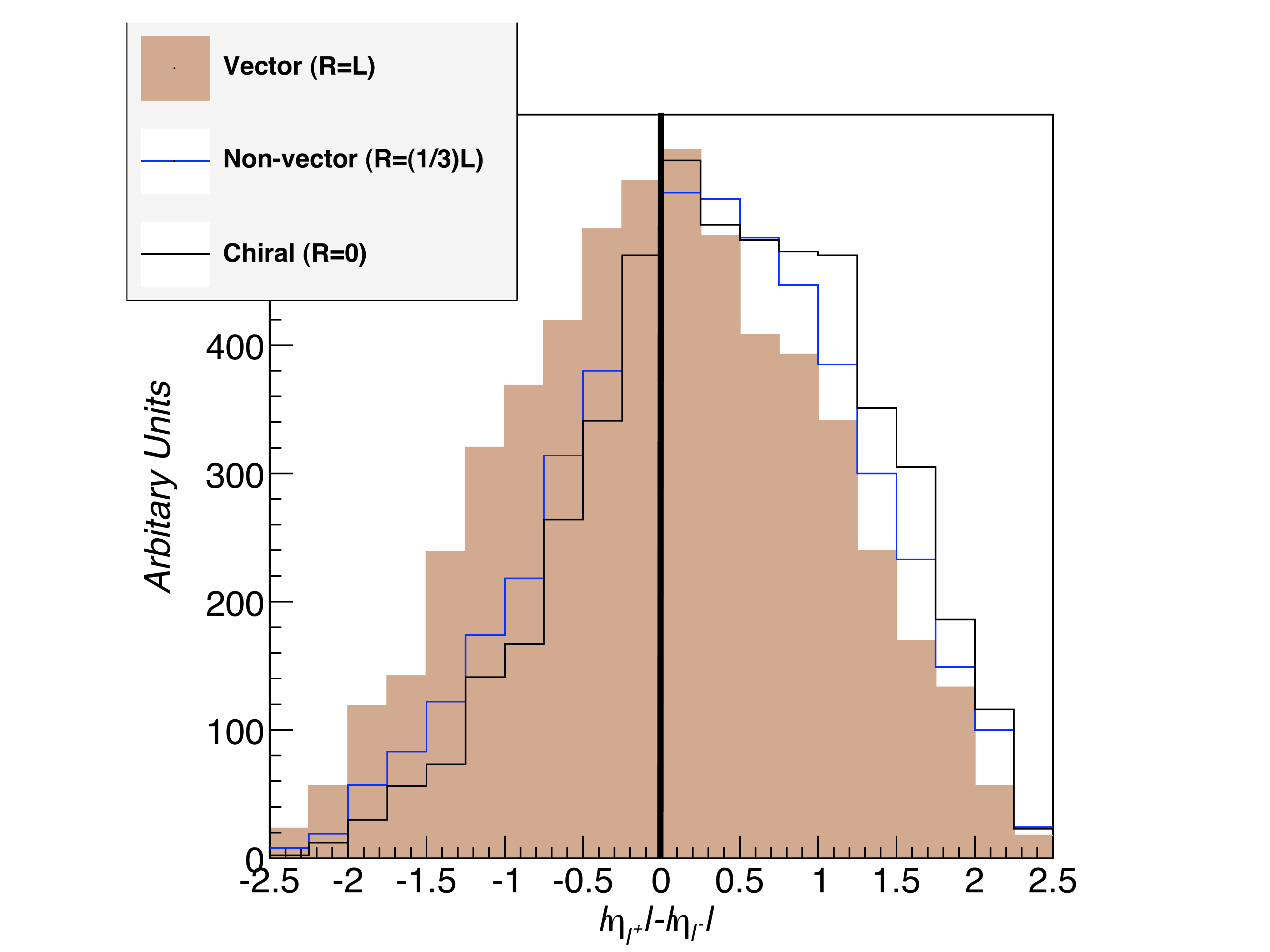}
\includegraphics[scale=0.2]{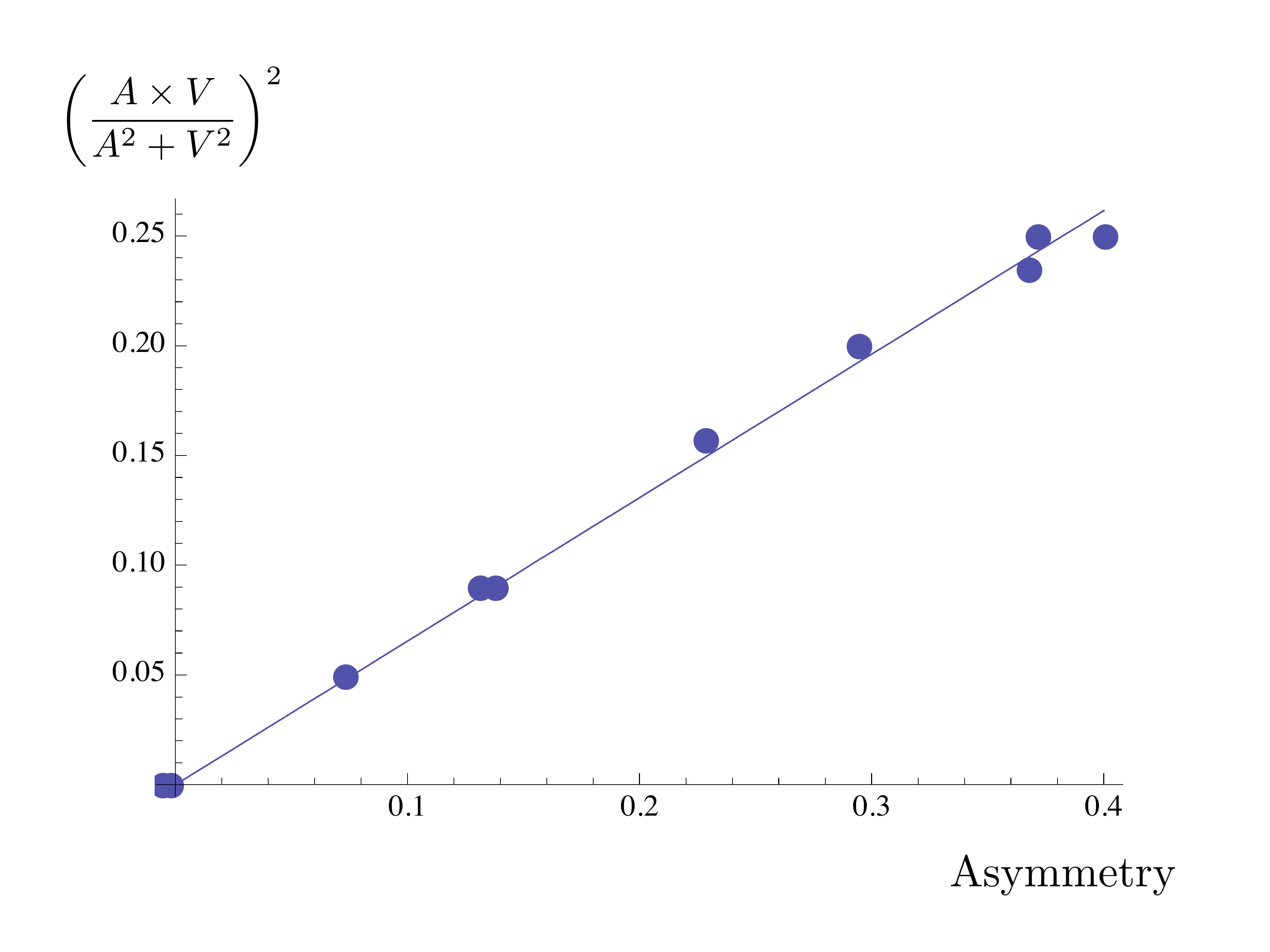}
\caption{Left: The distribution $|\eta_{\ell^+}|-|\eta_{\ell^-}|$ for different values of the resonance coupling to leptons and quarks. The solid distribution corresponds to a vector or axial coupling ($L=R$), and the blue and black distributions correspond to chiral ($L=0$ or $R=0$) and an intermediate case ($R=L/3$). Right: The value of the asymmetry $A_{\textrm{charge}}$ for several cases, and a fit to the coefficient in terms of $V$, $A$ couplings.} 
\label{asymmetry}
\end{figure}

First, note that the $Z'$ production is $q \bar{q}$ initiated, and there is no contamination from gluon-initiated processes as discussed in \cite{german}. But LHC is a $p \,  p$ collider, and identifying the direction of the incoming $q$ or $\bar{q}$ is not straightforward. Fortunately, $q$ is a valence parton at LHC, whereas $\bar{q}$ is a sea parton. When one convolutes the parton level  asymmetry in Eq.~\ref{partonAFB} with the distribution functions, one realizes that the quarks tend to have higher momentum than the anti-quarks. Therefore, the whole $q \bar{q}$ system  is boosted in the direction of the incoming quark.  We use this fact to obtain a charge asymmetry which is also proportional to the vector and axial couplings.

In this paper we propose a related, but different measurement. Instead of measuring the forward-backward asymmetry,  we define a charge asymmetry~\footnote{See Ref.~\cite{recentasym} for a recent discussion on the use of asymmetries on dileptons. Note that the asymmetry defined there differ from ours.}   
\bea
A_{\textrm{charge}} = \frac{N(\Delta \eta >0 - N(\Delta \eta <0 )}{N(\Delta \eta >0 +N(\Delta \eta <0 )}
\label{chasym}
\eea
where
\bea
\Delta \eta = |\eta_{\ell^+}|-|\eta_{\ell^-}|
\eea
Note that this asymmetry has been used before by the authors of Ref.~\cite{ross} to characterize s-channel resonances. The charge asymmetry is proportional to the asymmetry defined in Eq.~\ref{partonAFB}, and therefore provides information on the vector and axial couplings, as we will show below. The simulation in Fig.~\ref{asymmetry} confirms this expectation.
In Fig.~\ref{asymmetry} we plot the distribution of $\Delta \eta$ in the laboratory frame at 7 TeV for the decay of a 700 GeV resonance. The simulation of the asymmetry is done at parton level because the measured charge and $\eta$ values are close to the true value for  the electrons we have simulated, with cuts of  $p_T>$ 25 GeV and $|\eta|<$ 2.5.  Indeed, the electron charge misidentification rate is very low, of order $6 \times 10^{-3}$, and the $\eta$ resolution is of order $3\times 10^{-2}$~\cite{chargemisid}. Hence a full simulation is not necessary to get an accurate estimate of the charge asymmetry. 

The solid distribution corresponds to a vector or axial coupling ($L=R$), and the blue and black distributions correspond to chiral ($L=0$ or $R=0$) and an intermediate case ($R=L/3$). In the pure vector or axial case, the distribution is symmetric. We can also compute the total asymmetry and compare it with our theoretical expectation of the dependence with the couplings from Eq.~\ref{coeff}. The agreement is excellent and leads to a fit which accounts for all the parton distribution functions and the effect of cuts which are not encoded in the parton level Eq.~\ref{partonAFB}. For a 700 GeV resonance at 7 TeV COM energy, the fit leads to
\bea
A_{\textrm{charge}} \simeq 1.53 \, \left( \frac{A.V}{A^2+V^2} \right)^2 \ .
\eea
Once the mass of the resonance is obtained from the dilepton invariant mass spectrum, one can use a Monte Carlo simulation, the experimental value of the charge asymmetry and the dilepton rate (which is proportional to $V^2+A^2$) to obtain the couplings of the resonance to light fermions. 

Although the results in this section are model-independent,  let us mention the expectations for the three models considered in the previous section. CHL is the most predictive model in terms of couplings to fermions, as they have to be adjusted to suppress the $S$ parameter. In CHL, the measurement of the mass and rate in the dilepton channel can easily be inverted as a prediction for the couplings, and  hence checked against the measurement of the asymmetry.
In HTC, both resonances are an admixture of pure vector and axial, even before electroweak symmetry breaking, hence we expect $V\sim A$. In TCSM, the lightest resonance before electroweak symmetry breaking is a pure vector, but ends up varying from this expectation by 30$\%$ after the mixing with the SM gauge bosons.  

The charge asymmetry is a  measurement which could be done before the forward-backward asymmetry as it requires smaller statistics than a full differential angular distribution. The charge asymmetry and the forward-backward asymmetry are simply related because of angular momentum conservation. For two incoming RH (LH) particles, the initial state has +1 (-1) unit of angular momentum, while, as we are dealing with massless fermions, all amplitudes with an initial or final state with zero angular momentum are zero.  For a LH initial state, a LH final state will be produced with amplitude $(1+\cos{\theta})$ , where we denote $\theta$ as the angle between positively charged lepton and the beam axis; the amplitude is zero when the final angular momentum vector points opposite the initial. A RH initial state and RH final state gives the same distribution, while a LH (RH) initial and RH (LH) final state produces $(1- \cos{\theta})$. When a purely vector or purely axial particle is produced, all four combinations of helicity amplitudes contribute and sum to $\propto (1+ \cos^2{\theta})$. When a chiral (RH = 0 or LH = 0) resonance is produced, only one sub-amplitude contributes (LH $\rightarrow$ LH or RH $\rightarrow$ RH) and the distribution in the differential cross section is $(1+\cos{\theta})^2$. Therefore, the positively charged lepton from a chiral resonance sits preferentially at smaller scattering angle, or higher rapidity, while a vector/axial resonance produces forward/backwards leptons (or high/low rapidity) symmetrically. The preference for forward leptons in the chiral resonance case is what leads to the shift in $|\eta_{\ell^+}| -  |\eta_{\ell^-}|$. 

If the resonance couplings are purely vector or purely axial, then there is not information to be extracted from the charge asymmetry, but instead one has to resort to interference effects with the photon and $Z$ boson~\cite{interf}. The interference effect leads to a distribution
\bea
\frac{d \hat{\sigma}}{ d \cos \theta} \propto \textrm{ (Z' exchange) } + \left (V^2 \, (1+\cos^2 \theta)+ 2 A^2 \cos \theta \right) \, \Re \left( \frac{\hat{s}}{\hat{s}-m_{Z'}^2+i \Gamma_{Z'} m_{Z'}} \right)
\eea
where $\theta$ is the angle of the electron with the beam, see Ref.~\cite{interf} for details. But this measurement would require large statistics. Indeed, in an s-channel production of a resonance, $\hat{s}\sim m_{Z'}$, leading to a suppressed interference effect. 
  
\section{Conclusions}  

Scenarios of strong electroweak symmetry breaking are an attractive alternative to the fundamental Higgs mechanism. In these scenarios new resonances show up as composite objects of the strong dynamics. The common prediction to all these models is that the resonances would couple strongly to $W,Z$ bosons and help in the unitarization of $WW$ scattering. Unfortunately, a direct measurement of these couplings is difficult since it relies on the vector boson fusion channel, requiring a large luminosity and the capacity of forward jet tagging. 

We take a different approach in this paper. Even if the resonance couplings to light fermions is suppressed, the s-channel production may turn out to be the discovery channel. Resonances can be produced through quarks in the proton, and it opens the possibility of dilepton final states, which are very clean and provide charge correlations. 

Indeed, many scenarios predict sizable s-channel production. In this paper, we focus on three distinct scenarios based on warped extra-dimensions and technicolor scenarios, and use dileptons as the discovery and also characterization channel. The models considered here are Cured Higgsless (CHL), Holographic Technicolor (HTC) and Low-Scale Technicolor (LSTC), and we outlined their main characteristics in the text.

We first look at the dilepton invariant mass distribution. HTC has a very characteristic spectrum, with two nearby resonances. This degeneracy can be resolved experimentally, and it is a consequence of the viability of the scenario when confronted to precision tests. CHL displays two separated resonances with a ratio in mass fixed by the AdS geometry. LSTC just assumes a low-lying resonance, but makes no further assumptions. We have shown in this paper that one can distinguish between these spectra, which themselves imply very definite ways of addressing the electroweak precision constraints.

We then turn our attention to the charge information provided in the dilepton final state. We use this charge to construct a charge asymmetry which, combined with a rate in the dilepton channel, can be used to extract the chirality of the couplings of the resonance to the light fermions. This measurement would further help on setting apart different models, as it is related to the mixing of the resonance with the $Z$ boson and, in HTC, to the cancellation in the $S$ parameter. Moreover, the charge asymmetry is a  measurement which can be done before the forward-backward asymmetry as it requires smaller statistics. Again, this measurement yields information about the underlying structure of EWSB, and the way the model fulfills the requirements of electroweak precision measurements.

\section*{Acknowledgements}  
The authors would like to thank Travis Martin for pointing out a missing, and important, reference~\cite{ross}. 
P.B. would like to thank Aseshkrishna Datta, Y.K.~Wang, B.~Xiao and S-H.~Zhu
for useful discussions
AM is supported by Fermilab operated by Fermi Research Alliance, LLC under contract number DE-AC02-07CH11359 with the 
US Department of Energy. VS is partly supported by NSERC funding.

\end{document}